\documentclass[preprint,12pt]{elsarticle}

\usepackage{graphicx}

\usepackage{caption}

\usepackage{subcaption}

\usepackage{amssymb}

\journal{Astronomy and Computing}

\begin{document}

\begin{frontmatter}



\title{RFI Flagging Implications for Short-Duration Transients}


\author{
Y. Cendes,$^{1,2,3}$
P.~Prasad,$^{1}$
A.~Rowlinson,$^{1,4}$
R.A.M.J. Wijers,$^{1}$
J.D.~Swinbank,$^{5}$
C.J.~Law,$^{6}$
A.J.~van~der~Horst,$^{7}$
D.~Carbone,$^{1}$	
J.W.~Broderick,$^{4}$
T.D.~Staley,$^{8}$
A.J.~Stewart,$^{9}$
F.~Huizinga,$^{1}$
G.~Molenaar,$^{1,10}$
A.~Alexov,$^{11}$
M.E.~Bell,$^{12}$
T.~Coenen,$^{1}$
S.~Corbel,$^{13,14}$
J.~Eisl\"offel,$^{14}$
R.~Fender,$^{8}$
J.-M.~Grie{\ss}meier,$^{16,14}$
P.~Jonker,$^{17,18}$
M.~Kramer,$^{19,20}$
M.~Kuniyoshi,$^{21}$
M.~Pietka,$^{8}$
B.~Stappers,$^{20}$
M.~Wise,$^{4,1}$
P.~Zarka$^{22}$
\\
}

\address{
$^{1}$Anton Pannekoek Institute for Astronomy, University of Amsterdam, Science Park 904, 1098 XH Amsterdam, The Netherlands\\
$^{2}$Dunlap Institute for Astronomy and Astrophysics, University of Toronto, 55 St George St, Toronto, ON Canada\\
$^{3}$Leiden Observatory, PO Box 9513, 2300 RA Leiden, The Netherlands\\
$^{4}$ASTRON, the Netherlands Institute for Radio Astronomy, Postbus 2, 7990 AA Dwingeloo, The Netherlands\\
$^{5}$Department of Astrophysical Sciences, Princeton University, Princeton, NJ 08544, USA\\
$^{6}$Department of Astronomy and Radio Astronomy Lab, University of California, Berkeley, CA, USA\\
$^{7}$Department of Physics, The George Washington University, 725 21st Street NW, Washington, DC 20052, USA \\
$^{8}$Astrophysics, Department of Physics, University of Oxford, Keble Road, Oxford OX1 3RH, UK\\
$^{9}$Sydney Institute for Astronomy, School of Physics, The University of Sydney, NSW 2006, Australia\\
$^{10}$Department of Physics and Electronics, Rhodes University, PO Box 94, Grahamstown, 6140 South Africa \\
$^{11}$Space Telescope Science Institute, 3700 San Martin Dr, Baltimore, MD 21218\\
$^{12}$CSIRO Astronomy and Space Science, PO Box 76, Epping, NSW 1710, Australia\\
$^{13}$Laboratoire AIM (CEA/IRFU - CNRS/INSU - Universit\'e Paris Diderot), CEA DSM/IRFU/SAp, F-91191 Gif-sur-Yvette, France\\
$^{14}$Station de Radioastronomie de Nan\c{c}ay, Observatoire de Paris, CNRS/INSU, USR 704 - Univ. Orl\'eans, OSUC, 18330 Nan\c{c}ay, France\\
$^{15}$Th\"uringer Landessternwarte, Sternwarte 5, D-07778 Tautenburg, Germany\\
$^{16}$LPC2E - Universit\'{e} d'Orl\'{e}ans / CNRS\\
$^{17}$SRON, Netherlands Institute for Space Research, Sorbonnelaan 2, 3584 CA, Utrecht, the Netherlands\\
$^{18}$Department of Astrophysics/IMAPP, Radboud University Nijmegen, PO Box 9010, 6500 GL Nijmegen, The Netherlands \\
$^{19}$Max-Planck-Institut f\"{u}r Radioastronomie, Auf dem H\"{u}gel 69, 53121 Bonn, Germany\\
$^{20}$Jodrell Bank Centre for Astrophysics, School of Physics and Astronomy, The University of Manchester, Manchester M13 9PL, UK\\
$^{21}$NAOJ Chile Observatory, National Astronomical Observatory of Japan, 2-21-1 Osawa, Mitaka, Tokyo 181-8588, Japan\\
$^{22}$LESIA \& USN, Observatoire de Paris, CNRS, PSL/SU/UPMC/UPD/SPC, Place J. Janssen, 92195 Meudon, France}

\begin{abstract}
With their wide fields of view and often relatively long coverage of any position in the sky in imaging survey mode, modern radio telescopes provide a data stream that is naturally suited  to searching for rare transients. However, Radio Frequency  Interference (RFI) can show up in the data stream in similar ways to  such transients, and thus the normal pre-treatment of filtering RFI (flagging) may also remove astrophysical transients from the data stream before imaging.   In this paper we investigate how  standard flagging affects the detectability of such transients by examining the case of transient detection in an observing mode used for Low Frequency Array (LOFAR; \citep{LOFAR}) surveys. We quantify the fluence range of transients that would be detected, and the reduction of their SNR due to partial flagging.  We find that transients with a duration close to the integration sampling time, as well as bright transients with durations on the order of tens of seconds, are completely flagged.  For longer transients on the order of several tens of seconds to minutes, the flagging effects are not as severe, although part of the signal is lost.   For these transients, we present a modified flagging strategy which mitigates the effect of flagging on transient signals. We also present a script which uses the differences between the two strategies, and known differences between transient RFI and astrophysical transients, to notify the observer when a potential transient is in the data stream.
\end{abstract}

\begin{keyword}

radio transients \sep Radio Frequency Interference \sep RFI \sep Automated flagging \sep transient rates \sep transient rate limits



\end{keyword}

\end{frontmatter}


\section{Introduction}

Modern radio astronomy is particularly suited for the search of many classes of extreme transient object. This is due to a combination of astrophysical and instrumental effects. Astrophysically, extreme objects are often compact, non-thermal emitters, implying they can vary on short time scales and have significant luminosity at radio frequencies. At long radio wavelengths, the blackbody limit severely
constrains the allowed luminosity, but often compact objects are seen
to have coherent emission at those long wavelengths, which can exceed
the (incoherent) blackbody limit by many orders of magnitude.
Instrumentally, modern radio telescopes can have very wide fields of
view and thus are much more likely than previous instruments to detect rare events.
Further, modern computing techniques and hardware infrastructure enable the
near real-time processing of the large data streams from these
telescopes, and automated generation of alerts to transients.  This has enabled rapid multi-wavelength follow-up of candidate transients, essential for understanding the transient source parameters. However, the large data volumes lead to a requirement of automated flagging, which can be pernicious to transient searches.

There are multiple ways of using a radio interferometer to search for
transient events. One may use it to image the sky, and search for
variations of brightness in the sources detected in the image \citep{Stewart2016}. One may
also add up the signals from all the array elements coherently or
incoherently to form a beam on the sky and examine the time series of
the signal from that beam \citep{Coenen2014}. The latter method is preferred for looking
at very brief (sub-second) events and periodic signals, and is the
method of choice for detecting pulsars and Fast Radio Bursts (FRBs) \citep{Rane2017,Lorimer2007}. When pushed to
their limits, both methods require front-line data processing and
supercomputing technology in order to cope with the data flow \citep{LOFAR}. In
response to this, other techniques are now also being developed, such as
searching for suspected transient events in raw correlation data
products and then imaging very short selected data segments with
candidate transients \citep{Bower2013,vanVelzen2013}. 

In this paper we focus on the first method, searching for transients in a time series of images.  The effectiveness of image domain transient searches has already been demonstrated by the discovery of a bright, low-frequency transient of several minutes duration by \citet{Stewart2016}, as well as earlier discoveries by \citet{Hyman2009} and \citet{Jaeger2012}.  Interferometers spend a large fraction of their time collecting data in imaging mode, and at the same time sample visibilities initially at order of seconds intervals for the purposes of avoiding time smearing, rejecting bad data, and calibration.  This means that a large amount of data are collected in imaging mode that may be used to search for interesting transient sources, thus generating more science at no extra expense of observing time (but considerable extra processing time).

Typical image domain searches for transients of a given duration consist of tracking observations sampled at short timescales, followed by offline flagging and imaging at a much longer timescale in order to increase sensitivity. An example of this is the LOFAR survey mode, in which visibilities are integrated over 10 seconds in real-time, but imaging is carried out in 11 minute snapshots.

A key step in processing such radio data is the removal of Radio Frequency
Interference (RFI)-- terrestrial, generally human-generated, radio noise that
may corrupt celestial signals. In the large data volumes of e.g., LOFAR, automated algorithms
are needed to manage this process in realistic time. Generally these algorithms use the fact that
RFI is much stronger than cosmic signals in the data, and usually forms
stripes in a dynamic spectrum that are either narrow in frequency or in time,
on a smooth, much lower background. Simple removal of these data from consideration is adequate for most radio imaging, and even in a relatively RFI-rich environment like the Netherlands this results in only a few percent data loss \citep{Offringathesis}.

However, a bright and short astrophysical transient
may resemble RFI in the data, and thus be filtered out before any image
is created.  In this paper, we investigate the effect this has on the likelihood of transient detection.  We quantify the effect of various RFI flagging techniques on transients by injecting simulated transients into an actual LOFAR observation, then flagging and imaging the data.

The breakdown of the paper is as follows.  In Section 2, we will present an overview of current standard methods of RFI flagging, and of potential problems faced while flagging data from transient searches.  In Section 3, we will present our observations with the LOFAR Observatory and our transient simulations.  In Section 4, we will discuss the effect of various flagging strategies on simulated transients.  In Section 5, we will discuss the findings of our transient simulation results, along with a potential solution to identify astrophysical transients mistakenly flagged as RFI.  In Section 6, we will present our conclusions from this research.

\section{RFI Flagging Overview}

\subsection {Radio Frequency Interference and Celestial Radio Transients}

Radio Frequency Interference (RFI) from man-made sources is emission that can severely contaminate celestial observations by swamping the comparatively weak sky signal received by a
sensitive radio telescope.  The signature of received RFI in an observation depends on several factors,
including the type of radio telescope (single dish or interferometer), the type
of observation (continuum, spectral line, beam-formed or time resolved), and the kind of RFI
itself. The latter can range from impulse-like bursts, narrow-band, or wide-band
RFI, and can be ground-, air-, or space-based (e.g. terrestrial emitters, airplanes, spacecraft).

Temporally restricted RFI bursts are especially pernicious to transient
searches due to their resemblance to celestial transients. These can be both
band limited (e.g. Jovian bursts) as well as wide band. Thus, extra care needs
to be taken to distinguish such transients from RFI, and special RFI detection strategies
have been developed based on the class of transient \citep{Ryabov2004, Kocz2012}.


Interferometers used for imaging observations possess several natural defences
to RFI. A long enough spatial separation of the antennas making up a baseline results in a
different RFI environment at each antenna, leading to a natural suppression of
the RFI signal. Further, the delay tracking of a fixed location in the sky also leads to a
delay decorrelation of the RFI signal. The phase tracking of the interferometer
also leads to attenuation via fringe washing. Finally, interferometers usually
have high enough spatial resolution to be able to isolate a coherent RFI source in the
image plane.

In spite of these effects, interferometric observations can still be sensitivity
limited by residual RFI, due to the high sensitivity of the telescopes. Interferometers typically use offline post-correlation flagging methods, which can be applied over multiple passes on the data.  Of the available post-correlation methods, flagging using thresholding is by far the most common approach taken due to its ease of implementation and its predictable effect on the image noise, where the thresholding occurs on the visibility amplitude per baseline \citep{SumThreshold}.  Another approach uses filtering techniques, where a parametric model of the RFI signal can be built up and subsequently subtracted from the data, but this is computationally expensive and model dependent \citep{Offringa2012}. Adaptive interference cancellation using an independent estimate of the RFI signal obtained from reference channels also exist, but may not be as effective as the other methods \citep{Barnbaum1998}.  Spatial filtering can be an effective method, where the RFI source is isolated and subtracted from the visibility data \citep{Raza2002}. However, this is effective only if the RFI is significantly correlated at the antenna sites.  Higher order statistics can also be used for RFI flagging \citep{Fridman2001}, and Spectral Kurtosis in particular has been tested for transient detection \citep{Nita2016, Nita2010}.  However, the statistical methods such as Spectral Kurtosis require much more data processing in order to calculate the mean and variance, which is less feasible than thresholding methods for large-scale transient surveys.

For telescopes with large data volumes, such as LOFAR or the Murchison Widefield Array (MWA; \citet{MWA, Tingay2013}), automated algorithms are needed to manage this process in realistic time.  Generally these algorithms use the fact that RFI is much stronger than cosmic signals in the data, and usually forms stripes in a dynamic spectrum that are either narrow in frequency or in time,
on a smooth, much lower background. Simple removal of these data from consideration is adequate for most radio imaging.

Visibility based flagging would eliminate only the transients bright enough to cross a high threshold on a single visibility, leaving weaker transients relatively unaffected, and available in the image domain.  The latter could be many more in number, based on existing source counts \citep{Cordon1984}.  However, the current lack of a population of such transients and their unknown origins make the reliable and unambiguous detection of such events important, especially for triggering multi-wavelength observations. Detecting the brightest transients is thus essential in the discovery phase.

\begin{figure}
\label{fig:rfi}
\centering
\includegraphics[width=1\textwidth]{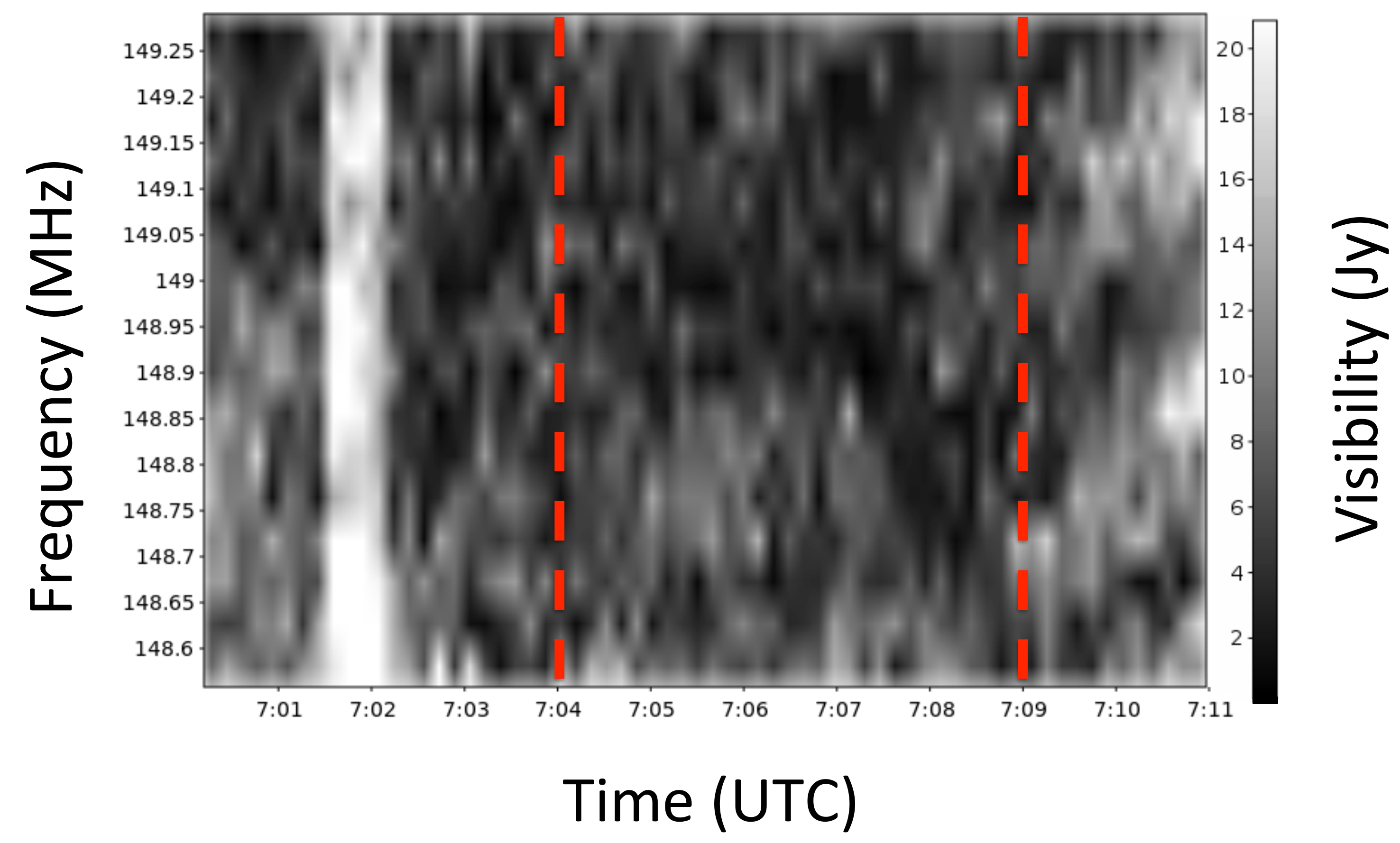}
\centering
\includegraphics[width=1\textwidth]{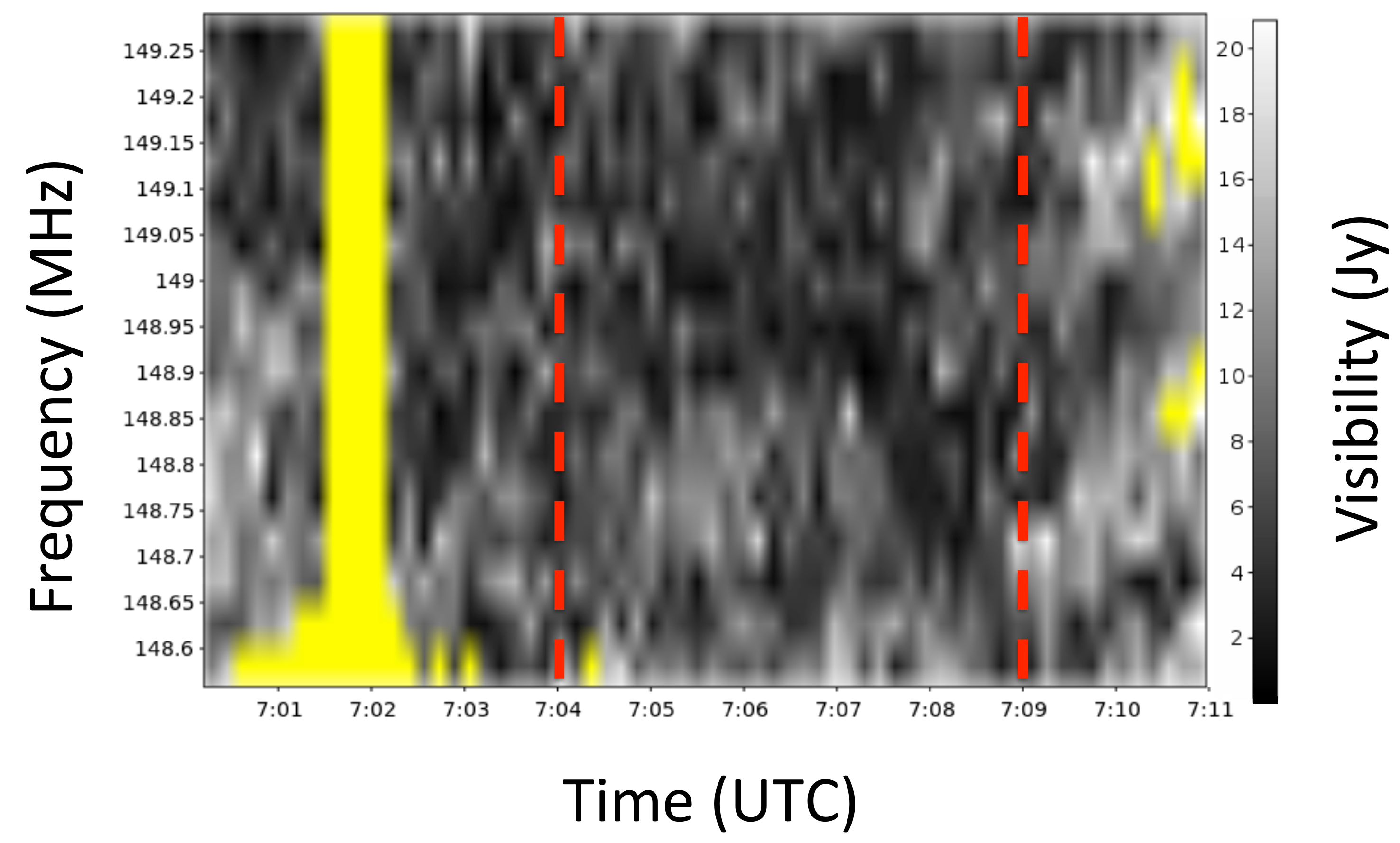}
\centering
\caption{A visual representation of the observed visibility amplitude in the time-frequency domain on the baseline between two LOFAR stations in the visualization program ``rfigui," where the top is an unflagged data set and the lower left has been flagged by AOFlagger (yellow).  Here we have an 11 minute LOFAR data set with a 30 second top hat simulated signal of amplitude 10 Jy injected at 07:01:30.  This signal was flagged as RFI when the flagging algorithm is run with default parameters, showing the potential perils of using automatic flagging during transient searches.  For this baseline, the 10 Jy signal is added on top of the median amplitude values.  During the period where no transient is present, indicated as the space between the red dotted lines, when looking at the amplitudes of the visibility time series the data has a standard deviation of $\sim$2.6 Jy and a median value of 5.5 Jy.  Here the scale is 20 Jy due to bright amplitude values at the end of a few channels, which are later flagged.}
\end{figure}

An illustration of the effects of thresholding based flagging on a bright transient can be seen in Figure 1, where the visibility amplitudes on a single baseline are visualized in the time-frequency domain. A simulated transient  of 30 seconds duration has been injected into the visibilities.  Through the sequence described in more detail below, the flagger will flag such a transient automatically as contaminated data.  As such an example illustrates, it is crucial to understand whether such transient signals spanning a range of amplitudes  could potentially be flagged before the data are even examined by the observer.  This aim is also important in light of transients such as the one reported by \citet{Stewart2016}, and surveys for FRBs using the MWA such as \citet{Tingay2015} and \citet{Rowlinson2016}, which all relied on the RFI flagging algorithm, AOFlagger \citep{AOFlagger}, for their analysis. We analyze the performance of this algorithm via simulations.

\subsection{RFI Flagging Algorithms}

\subsection{AOFlagger}

The current default RFI flagging software for radio imaging at multiple observatories, which is used routinely during transient surveys such as MWA \citep{OffringaMWA} and LOFAR, is called the AOFlagger.  In recent years, AOFlagger has become one of the standard RFI flagging algorithms, and is increasingly used on many radio telescopes.  It has been successfully used with data from several interferometric arrays, such as Very Large Array \citep[VLA;][]{Williams2017}, Westerbork Synthesis Radio Telescope \citep[WRST;][]{Adebahr2017}, the Australia Telescope Compact Array \citep[ATCA;][]{ATCA2016}, the Large-Aperture Experiment to Detect the Dark Ages \citep[LEDA;][]{Price2017}, the  Boolardy Engineering Test Array of the Australian Square Kilometre Array
Pathfinder \citep[ASKAP BETA;][]{Allison2017}.  It has also been successfully implemented on single dish telescopes, such as Parkes \citep{Offringa2012} and Arecibo. \footnote[1]{A full list of telescopes which have successfully used AOFlagger, along with the software itself, is available at: https://sourceforge.net/projects/aoflagger/}

In brief, the AOFlagger is an iterative algorithm which operates on a frequency resolved timeseries of visibilities from a single baseline, and can operate on either a single polarization or a chosen Stokes parameter. It fits a surface to the visibility amplitudes in the time-frequency plane in order to eliminate systematics, e.g., due to fringes from bright sources in sidelobes, and then clips out visibilities with amplitudes crossing a chosen threshold. The flagged visibilities are ignored in the next iteration. The algorithm detects low level RFI using a set of decreasing thresholds, implemented via the {\sc SumThreshold} algorithm.  It has been empirically determined that two iterations of {\sc SumThreshold} are adequate for flagging \citep{LOFAR-AOFlagger}.  The algorithm consists of the following steps \citep{LOFAR-AOFlagger}:

\begin{itemize}
\item \textbf{{\sc SumThreshold}}: This is a combinatorial thresholding algorithm where the same dataset is subjected to a set of decreasing thresholds. A threshold $\chi_{M}$ is determined by $M$, the number of samples in the data set surrounding a candidate data point (visibility amplitude for one time/frequency sample), where $M$ is predefined, and a chosen false detection rate \citep{SumThreshold}. If the sum of amplitudes of $M$ contiguous visibilities exceeds the threshold $\chi_{M}$, then all $M$ visibilities are flagged.

    The algorithm processes data in order of decreasing thresholds $\chi_{1}, \chi_{2}, ..., \chi_{M}$, where the thresholds are in units of the visibility amplitude standard deviation, $\sigma$.  The single sample threshold, $\chi_{1}$ is the highest, and eliminates really bright and sporadic visibilities. Visibilities flagged by a  $\chi_{i}$ threshold are ignored when computing the statistics for applying the next level threshold $\chi_{i+1}$.  These lowered thresholds are used to eliminate low-lying RFI not filtered by $\chi_{i}$.  Visibilities comprising such low-level RFI are usually connected to their neighbors in time and frequency, i.e., the connected visibilities form islands of high amplitudes. In the {\sc SumThreshold} algorithm, the range of applied thresholds depends on the extent of connectivity in all four directions on the time/ frequency visibility amplitude plane.

For a given baseline, {\sc SumThreshold} is applied twice in an AOFlagger iteration, first prior to the surface fit in order to ignore RFI when fitting, and then again when the fitting has converged in order to establish the actual flags.  The $\chi_{i}$ thresholds are decreased exponentially in every iteration of {\sc SumThreshold} to increase the sensitivity to low level RFI.

\item \textbf{Channel and Time Selection:} Next there is a channel and time selection step which flags problematic channels and time steps which may be fully contaminated but have not yet been flagged.  In the case of channel selection, the RMS value of visibilities across channels for a given time slice is computed.  The time series of these RMS values are Gaussian smoothed, and the standard deviation, $\sigma$, of the difference between the RMS time series and the smoothed time series is computed.  Any RMS value found to be $> 3.5 \sigma$ results in the entire set of channels for that time slice to be flagged.  The same occurs for the time axis. This step, akin to conventional thresholding, is implemented in order to facilitate faster convergence.

\item \textbf{Surface fitting:} Surface fitting then occurs, which removes fringes caused by strong sources in the side lobes in order to increase accuracy. This is done with a Gaussian kernel sliding window in both time and frequency space, and is a time-consuming step when compared to the other steps.

\end{itemize}

Overall, for routine data collection and subsequent imaging, the AOFlagger has proven to be a fast and accurate automatic flagger, particularly in the relatively high-RFI LOFAR radio environment \citep{LOFAR-RFI}, although it can, in some cases, flag data that is not RFI \citep{Offringathesis}. Normally, however, this is not an issue in routine observatory observations because mistakenly flagged data are a small percentage of the entire data set \citep{LOFAR-RFI,OffringaMWA}.

\begin{figure}
\includegraphics[width=0.8\textwidth]{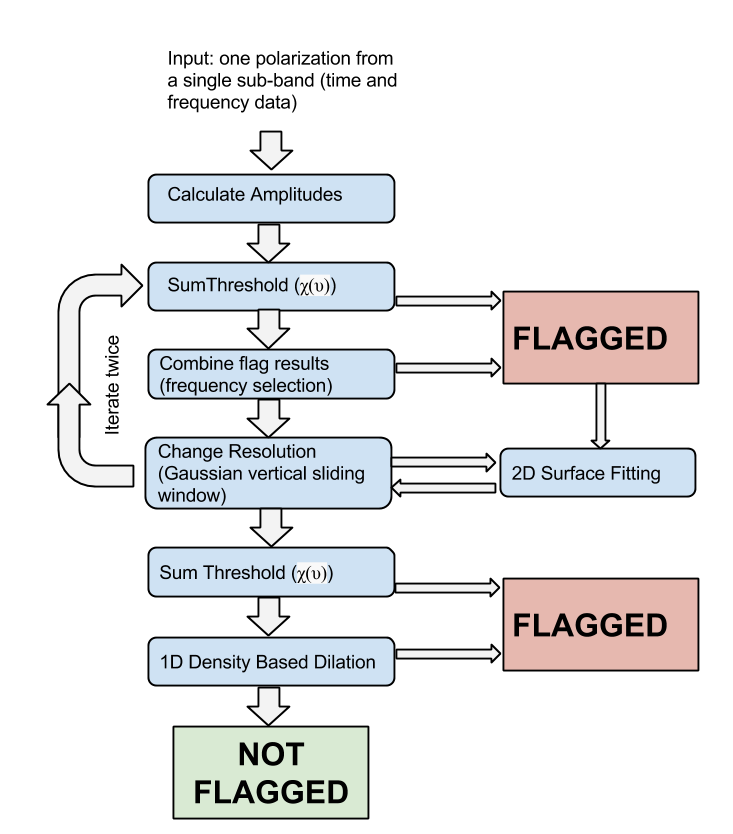}
\centering
\caption{A flow chart of the steps in AOFlagger.  The steps that we focused on in particular for our transient specific analysis for modification are the {\sc SumThreshold} and time selection steps.  We should note the input can also be a Stokes parameter of interest.}
\label{fig:flow-chart}
\end{figure}

A flow chart of AOFlagger can be seen in Figure \ref{fig:flow-chart}.

\subsection{Concerns for Transient Searches}

The aim of this paper is to investigate the effect of AOFlagger on transient observations.  As such, in addition to investigating the impact of the default AOFlagger settings on transients, we also focused on modifying the flagger to see whether an automatic RFI flagging method could be adapted for the needs of transient searches.  In particular we examined the following:

\begin{itemize}

\item One particular concern was that the thresholding algorithm, {\sc SumThreshold}, works in both the time and frequency domains.  This means brief, bright transients that are broadband, and a focus for image-plane surveys, could be flagged due to their sudden increase of power in the time domain.

\item Time selection steps that rely on sigma clipping may inaccurately flag data that are not RFI.  For example, a sudden, brief transient in an otherwise quiet radio field would have its amplitude suddenly change by a high value compared to the data around it, and it could be mistaken for RFI and flagged at this stage.

\item In addition to this, the lowered thresholds in later steps of {\sc SumThreshold} for data connected in either frequency or time are also a concern for transient searches.  This is because such lower thresholds may flag fainter transients which are normally not affected by the high single sample threshold $\chi_{1}$.

\end{itemize}

\section{Observations and Transient Simulations}

\subsection{Observations}
\label{sec:observations}

As we were interested in examining the consequences of RFI flagging on real data, we wanted to examine a field observed with LOFAR.  For this purpose, we chose data taken with the LOFAR telescope as an extra target observation during routine transient monitoring observations (Fender et al., in prep) using one sub-array pointing of the telescope, where the beam is formed using the innermost antennas from each station.  We then focused on a low-RFI 11 minute measurement set, taken on February 10, 2013, using one beam at 149 MHz with a bandwidth of 781 kHz, along with calibrator source Cygnus A. 



First, the data were pre-processed using the standard methods with LOFAR: they were flagged for RFI using AOFlagger \citep{AOFlagger}, and the data calibration and imaging were carried out using the method described by \citet{Broderick2016}.  We used a model of the source for the calibration of the calibrator sub-bands, and then the gain amplitudes and phases were transferred to the target field data as outlined by \citet{Heald2011}.  Specifically, we used a model of Cygnus A from \citet{McKean2011}. We also refined the calibration further by performing phase-only self-calibration on the target field, as outlined by \citet{LOFAR}.

For imaging, we used the AWImager \citep{Tasse2013}; a maximum projected baseline of 6 km was chosen to ensure a reliable image from the given u,v coverage
for the 11 min snapshot, leading to an angular resolution of approximately 34 arcsec. This image has a
total field of view of 11.35 deg$^2$ in a circular region.  We measure a noise standard deviation of  25 mJy by generating a noise  map of the image using the sourcefinding script PySE (Carbone et al., submitted)


\subsection {Observational Detection Bounds on Transient Fluence}
\label{sec:test-case}
We quantified the effect of a threshold based flagging strategy on a bright transient by simulating the presence of such a transient on a dataset with a realistic noise and RFI instantiation.  Assuming no sources in the field, the measured LOFAR System Equivalent Flux Density (SEFD) for a core station in HBA mode is $~3 $ kJy at 150 MHz \citep{LOFAR}.  To calculate the expected noise on a single visibility amplitude of a single polarization, $\Delta S$, under the assumption of a high SNR, we use the equation 

\begin{equation}
\Delta S = \frac{1}{\eta}\frac{SEFD}{\sqrt{2}\sqrt{\delta\nu t_{int}}}
\label{eq:noise}
\end{equation}  

where $\delta\nu$ is the channel width, and $t_{int}$ is the integration time.  For our data, the system efficiency $\eta$ is assumed to be 1, and the two stations making up the baseline are assumed to have the same SEFD.  The subband of 781 kHz was spectrally resolved into 16 channels, with visibility integration times of 10 seconds. This corresponds to a noise $\Delta S$ of 3.04 Jy per visibility of a single polarization. Thus, if the time selection threshold is set at its hard-coded default as 3.5$\sigma$, we can calculate that any transient signal exceeding $S_{flag} = 10.6$ Jy will be flagged for these observing parameters.

For determining the theoretical flagging bounds, we used the single sample threshold due to its effect on short duration (1 sample) transients. At the single sample level, two thresholds are applied to the data by AOFlagger: $\chi_{1}$, and the time selection threshold of $3.5\sigma$. We calculate the flagging bound based on the time selection threshold due to it being the more conservative (lower) of the two.

We estimated the standard deviation of the visibility amplitude over the 11 minute observation for a typical baseline, and this was found to be $\sim$2.5 Jy.  We chose to carry out our analysis with the per visibility noise derived from the empirical SEFD due to the expectation of LOFAR sensitivity to be of that order, and the fact that there can be systematics in the visibility time series such as fringes due to bright sources which can bias the measured statistics.

When flagging is applied on high time and frequency resolution visibilities and then transient detection is carried out on the integrated image, transients can be lost due to two reasons:
\begin{itemize}
 \item The transient flux per visibility exceeds the used flagging threshold, or
 \item The transient fluence (flux$\times$duration) is lower than the image detection threshold.
\end{itemize}
Thus, in order to be detected via imaging, transients need to lie in the region where they are bright enough to be visible in the integrated image, but not bright enough to be flagged.

\subsubsection {Effect of transient duration on its detection}
We can compute the duration a transient should have to be visible in an integrated snapshot by limiting the transient flux to be $< 10.6$ Jy to prevent its visibility flagging.  Due to computational limitations, typical searches for these brief signals in the image domain will be in integrated images, such as the 11 minute observations of LOFAR.  As outlined by \citet{Trott2013}, if an image is longer than the width of the transient signal duration, then we can estimate the minimum transient signal flux density, $S_{min, w}$, that LOFAR is sensitive to using:

\begin{equation}
S_{min, w} = S_{min, 11min}\frac{\Delta t}{w}
\label{eq:dilution}
\end{equation}  

where $S_{min, 11min}$ is the sensitivity of one snapshot image multiplied by the detection threshold, $\Delta t= 660$ s is the snapshot integration time, and $w$ is the duration of the transient.  The background noise of the 11 minute integrated image is 25 mJy beam$^{-1}$, as measured by our observations, and we set a $5\sigma$ detection limit for a source in the image plane, corresponding to a source flux of $S_{min, 11min} = 0.125$ Jy in an image with $\Delta t = 660$ s.  From Equation \ref{eq:noise}, we showed sources with a flux greater than 10.6 Jy will be flagged.  Therefore, using Equation \ref{eq:dilution}, we can show that transients with a duration $w \lesssim 8$ s would not be detected due to the channel selection step, although transients of longer duration could also be flagged due to other parts of AOFlagger such as {\sc SumThreshold} or surface fitting.  

\subsubsection {Fraction of transients lost}

We can quantify the fraction of transients lost due to flagging from time selection and fluence by assuming that the number density of transient sources is constant in a Euclidean universe. Therefore, knowing the  cumulative flux distribution, $log(N)-log(S)$, of the sources, and assuming a simple power law, $N (>S) \propto S^{-\alpha}$, we can determine the fraction of sources expected to be lost due to flagging via:

\begin{equation}
\frac{N(S > S_{flag})}{N(S > S_{min, 11min})} = \frac{S_{flag}^{-\alpha}}{(S_{min, 11min}\frac{\Delta t}{w})^{-\alpha}}
\label{eq:alpha}
\end{equation}  

For $w < 8$ s, all transients will be lost.  For longer durations, let us take the concrete example of a Euclidean universe, where transient sources are evenly distributed and radio source count for transient sources brighter than a given flux would follow a cumulative distribution for logN-logS $\alpha = 1.5$.  For a transient which lasts 5\% of the measurement set, or 33 seconds, $\sim11$\% of transients detectable in the image will be flagged based on Equation \ref{eq:alpha}.  These transients are significant because they give the highest S/N, and are the best candidates for multi-wavelength follow up and understanding the physics behind the source population.

%
%
%

The fraction of transients expected to be rejected via a flagging operation is dependent upon the transient duration and the extent of the imaging dwell time.  It increases as the transient duration reduces to approach the hard limit of 8 seconds, after which all transients are undetectable.  The fraction flagged then decreases as the transient duration increases to that of the snapshot integration. With theoretical boundaries set up, we populated the range of fluences expected to be affected by a default flagging strategy in order to quantify the effect of flagging on transient detection.

\subsection{Transient Simulations}
\label{sec:simulations}

In order to quantify the effect of a flagging algorithm based on applying a set of thresholds on the visibility amplitude, we added simulated transients onto an actual LOFAR observation which would have a realistic noise and RFI instantiation. We focused our efforts on an observation at 149 MHz consisting of a single snapshot of 11 minutes and with a total bandwidth of 781 kHz, where the visibilities were already calibrated.  These data were available at a spectral resolution of 48.8 kHz and a temporal resolution of 10 seconds.  As described earlier, data were already flagged once with AOFlagger, and the visibilities calibrated.  The tests we describe here involve a secondary RFI flagging step after the transient was simulated and added to the calibrated visibilities, but before imaging. This was done in order to study the effect of the flagger on this one injected transient signal in particular, as this data was very low in the amount of RFI it contained.

For these simulations a transient source was injected in the center of the image by taking a model point source, transforming it to visibility space, and adding this to the recorded visibilities.  This was done using BlackBoard Selfcal (BBS) software \citep{Loose2008} where the temporal profile of the point source was a top hat signal of a given strength and duration in the image's sky model.  After this injection, the data were processed as usual through the imaging pipeline described in Section \ref{sec:observations}, and the resulting image was then analyzed.

This transient’s amplitude was varied from 0.5-10 Janskys in half Jansky intervals, and the duration was varied from one second to 2 minutes on a logarithmic scale. This range, based on estimates laid out in Section \ref{sec:test-case}, was selected to probe the range of $\chi$ thresholds which depend on both the amplitude and connectivity of visibilities. The resulting images for these data were then run through the sourcefinding script PySE where the integrated flux at the location of the injected transient was measured via a forced fit of a point source.  We should note that the data we injected the transient into had already been run through RFI processing during pre-processing when it was first obtained for analysis– that is, the typical RFI processing for LOFAR imaging. The tests we describe here involve a secondary RFI flagging step after the transient was simulated and injected into the data but before imaging in order to study the effect of the flagger on this one injected transient signal in particular, especially as these data contained a very low amount of RFI.

The results of this test before any secondary flagging steps were introduced can be seen in Figure \ref{fig:11min-unflagged}.  Here, we see the brightest detected transients correspond with the brightest simulated transients with the longest duration, and the measured flux of the transient (indicated by the color bar) is consistent with  expectations  from Equation \ref{eq:dilution}.  This measured flux then decreases for transients of shorter durations and brightnesses, until they are undetectable because they are either too faint, too brief, or a combination of the two.  After this, the same tests were conducted but with an automatic flagging step added after the transient was injected into the data in order to measure the effects of RFI flagging on the transient signal.  

We also performed a similar simulation for radio transients of varying brightness but with 2 minute long snapshots, which was the same duration as the \citet{Rowlinson2016} MWA survey for brief transients, flagged and calibrated using 2 minute measurement sets but imaged in 30 second increments.  This was done in order to understand how flagging for transients would affect shorter measurement sets, because the number of data samples is much smaller for such observations.  For these measurement sets, we focused on transients that were up to 60 seconds duration in the data, or half the length of the observation.  Further, we note that because AOFlagger has a recommendation of at least 1,000 time steps to be passed through the flagger, and this is impossible for MWA-length measurement sets, we edited the time window in which the measurement sets are divided for flagging into smaller increments than is recommended.

\section{Effect of Flagging Strategies on Simulated Transients}

\subsection{Default AOFlagger Settings}
\label{sec:default}

\begin{figure}
\quad
\begin{subfigure}[b]{.5\columnwidth}
\centering
\includegraphics[width=1\textwidth]{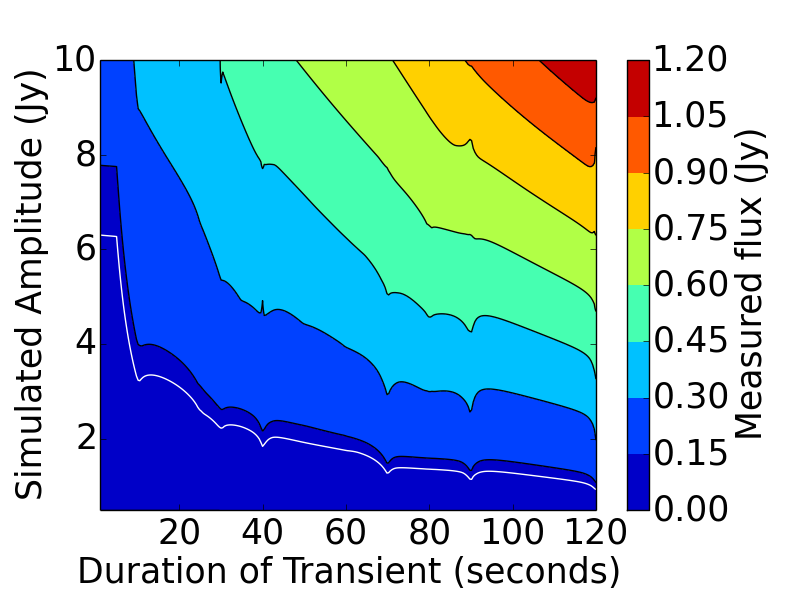}
\caption{11 minute measurement set with unflagged data.}
\label{fig:11min-unflagged}
\end{subfigure}
\quad
\begin{subfigure}[b]{.5\columnwidth}
\centering
\includegraphics[width=1\textwidth]{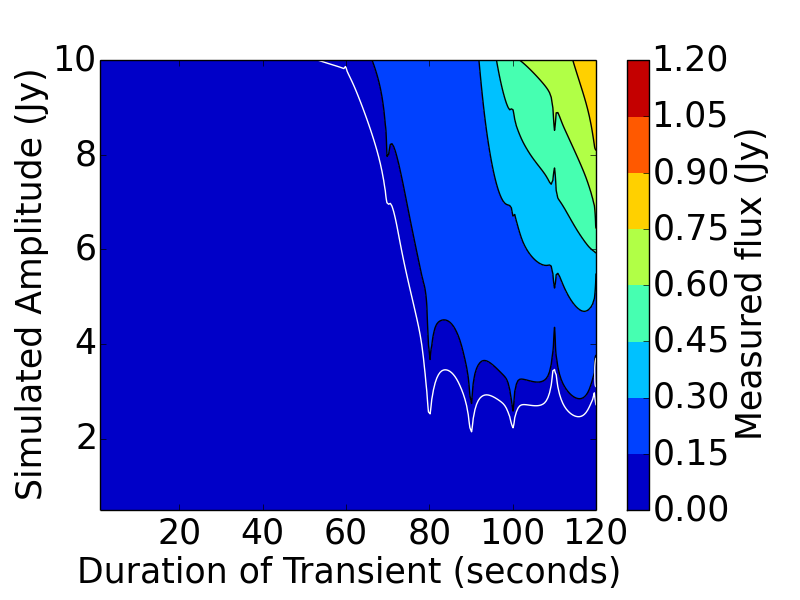}
\caption{11 minute flagged with the default flagger.}
\label{fig:11min-default}
\end{subfigure}
\quad
\begin{subfigure}[b]{.5\columnwidth}
\centering
\includegraphics[width=1\textwidth]{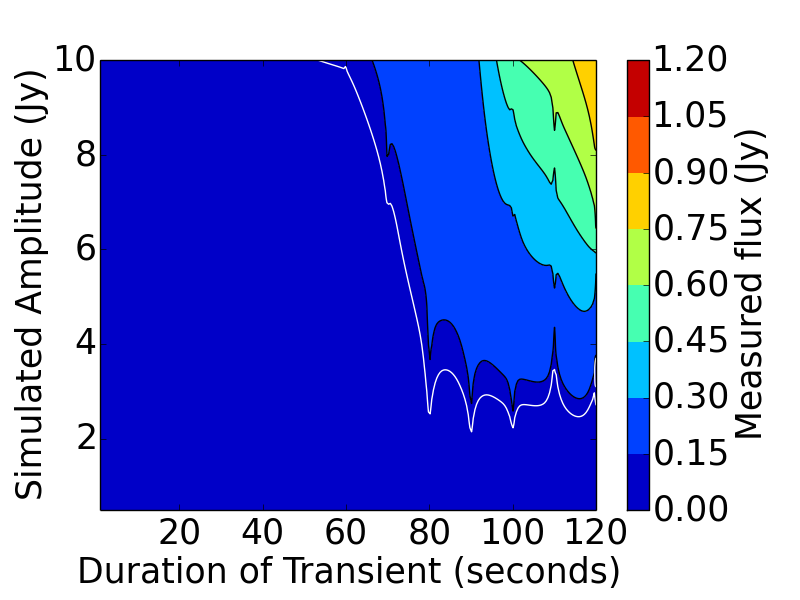}
\caption{11 minute flagged data with 6.5$\sigma$ time selection.}
\label{fig:11min-6sig}
\end{subfigure}
\quad
\begin{subfigure}[b]{.5\columnwidth}
\centering
\includegraphics[width=1\textwidth]{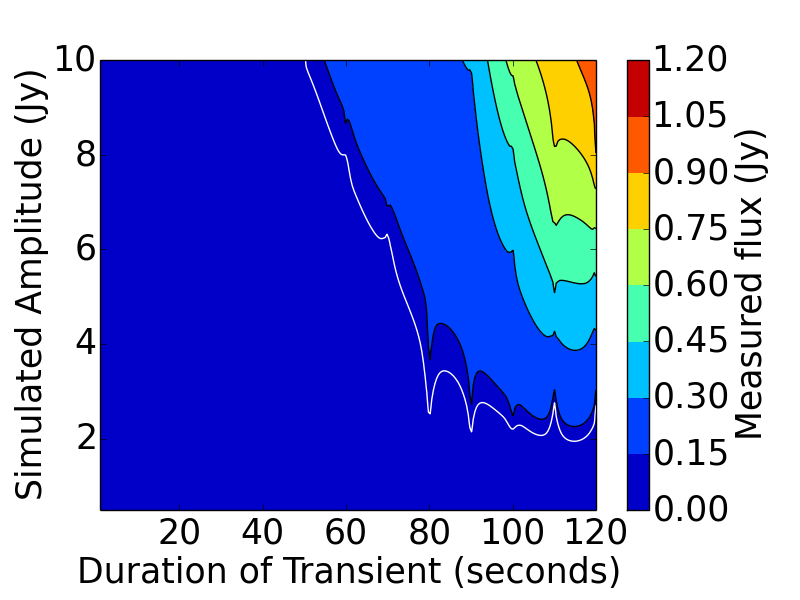}
\caption{11 minute flagged data, no time selection.}
\label{fig:11min-output}
\end{subfigure}
\quad
\caption{Observed fluxes for transient signals of a given amplitude and duration in an 11 minute measurement set.  Here, the top left plot is with unflagged data, the top right plot is for the default flagger, the bottom left is with the modified flagger with {\sc SumThreshold} only in the time direction but a time selection $\sigma >$ 6.5, and the bottom right is with {\sc SumThreshold} only in the time direction and with no time selection steps.  The unusual contours in the flagged data between levels is due to a sampling effect caused by the sampling window size of the RFI algorithm and brightness of the transient.  The white contour corresponds with $S_{min, 11min} = 0.125$ Jy, which is the $5\sigma$ limit derived from the RMS map of the image.}
  \label{fig:11minfluxes}
\end{figure}

We first considered only the effects of  the standard AOFlagger settings on the flagging of transients, in order to characterize the typical effects of such flagging on bright transients.  In an 11 minute measurement set, we found that if the simulated transient was of a longer duration than two minutes there was no statistically significant difference in the observed flux of the transient\textemdash that is, the transient was unaffected by flagging algorithms because its long duration would not trigger the thresholding algorithm.  However, when considering brief transients of durations less than two minutes, as seen in Figure \ref{fig:11minfluxes}, differences between the two become apparent.  In Figure \ref{fig:11min-unflagged}, where no flagging occurs, the transients with high simulated amplitudes correspond with significantly higher measured fluxes than in Figure \ref{fig:11min-default}, which had the default AOFlagger settings run on it before fluxes were measured.  If automated flagging is used, very brief transients ($\sim\Delta t < 60$ s) of a detectable brightness will be flagged out altogether by the automated flagging software because of the single sample threshold, $\chi_{1}$, designed to eliminate very bright and sporadic RFI.

It should be emphasized that the transient signal measured at the end of this procedure would be weaker than the original signal injected, as described by Equation \ref{eq:dilution}.  Thus, in Figure \ref{fig:11min-unflagged}, for example, the transient signal flux measured is less than the amount injected.

\begin{figure}
\centering
\includegraphics[width=1\textwidth]{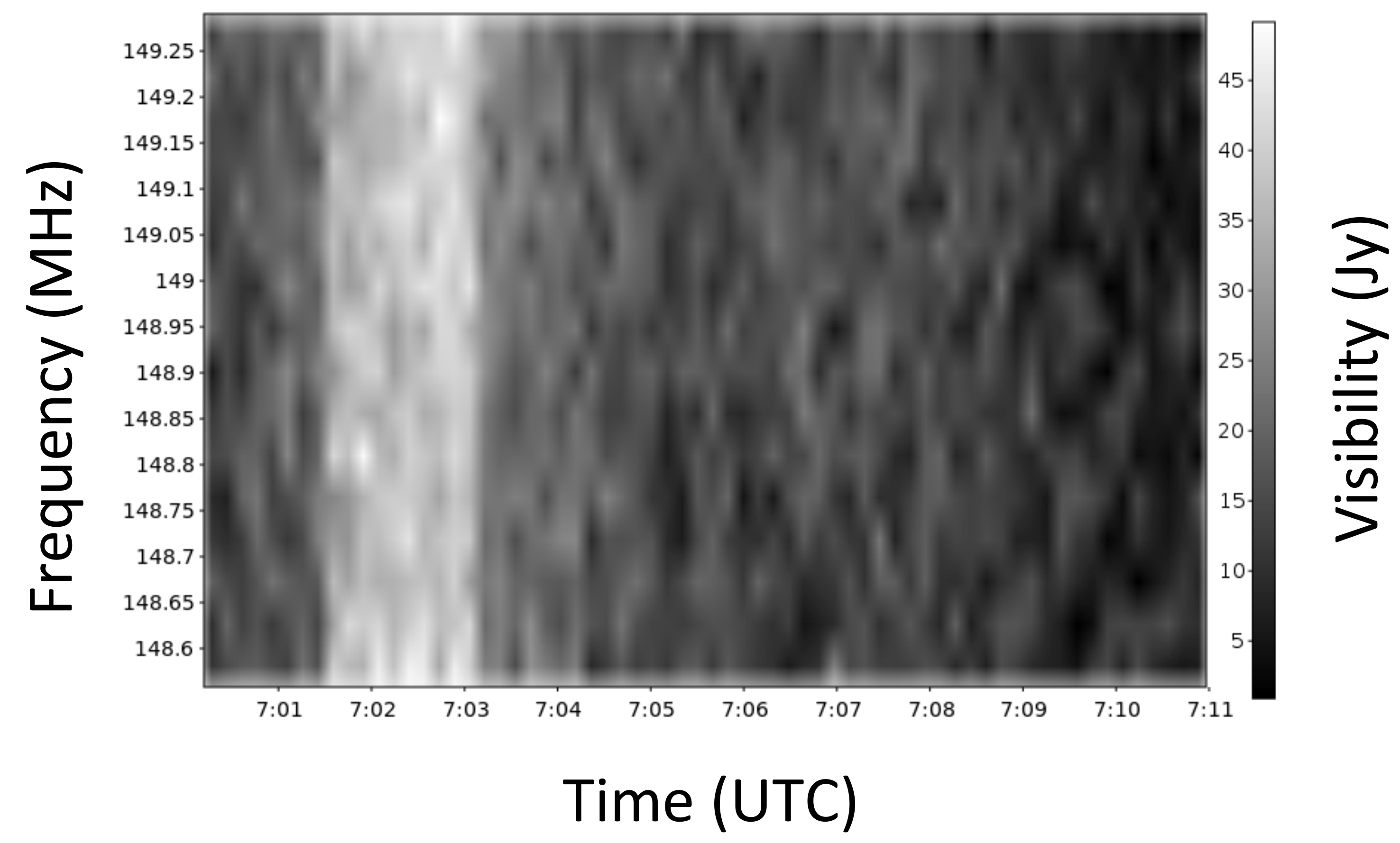}
\centering
\includegraphics[width=1\textwidth]{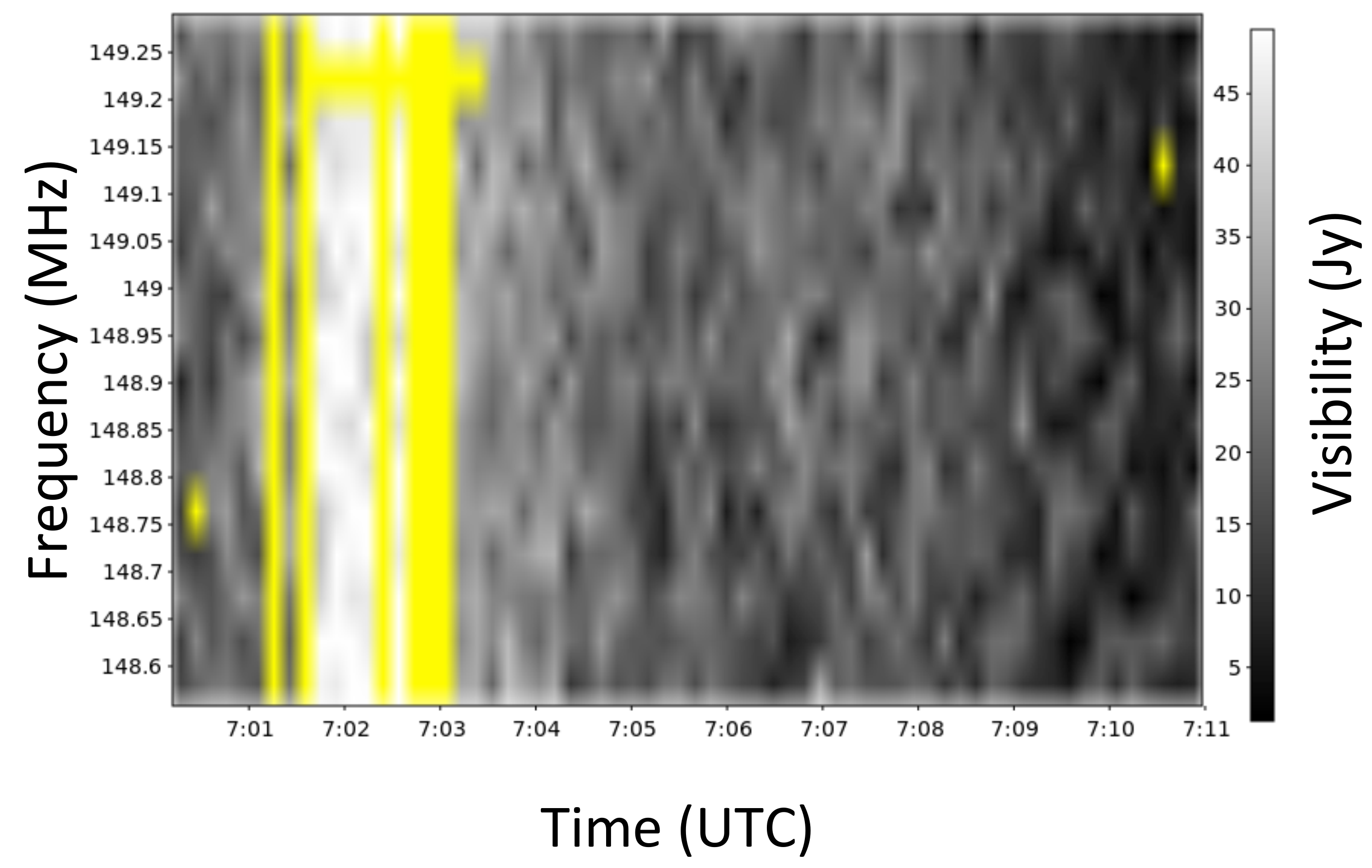}
\centering
\caption{An illustration of "partial flagging" in the time-frequency domain for two antennas from the program "rfigui," where the top is an unflagged data set and the bottom has been flagged by AOFlagger (yellow).  Here we have an 11 minute data set with a top hat transient from an injected signal with 10 Jy amplitude and 90 seconds duration.  A sizeable percentage of the transient is flagged as RFI but not the entire signal.  As such, the transient would be detected, but not at its full brightness or duration.  We note that some data around the transient which does not contain the injected signal are also flagged.}
\label{fig:partial-flagging-example}
\end{figure}

Longer, brighter transients (60 s $< \Delta t < $120 s), however, are no longer flagged completely by AOFlagger, but do appear to have a lower flux when compared to the original strength of the injected transient.  A check of what is being flagged in the visibilities reveals that the transient signals are becoming "partially flagged," where some, but not all, of the signal is being flagged as RFI.  An example of such partial flagging can be seen in Figure \ref{fig:partial-flagging-example}.  Exactly how much of the signal is being flagged is dependent on the intensity and duration of the signal.  We also note that sections of the non-transient data are also being flagged by the flagger in order to eliminate any potential fringes of RFI within the data due to dilation, which can further affect the measured fluxes in the data stream.  For example, for the unflagged data, the 10 Jy transient with 2 minutes duration had a measured flux of 1.24 Jy (Figure \ref{fig:11min-unflagged}).  For the flagged data, the same transient was measured as 1.44 Jy in brightness (Figure\ref{fig:11min-default}.  This was because while $\sim30$\% of the transient signal was flagged, a large fraction of data surrounding the signal which did not contain a transient was also flagged by AOFlagger, making the effective $\Delta t$ of the measurement set shorter than for the unflagged data.  This can have a significant effect: for example, a 4 Jy transient of 100 seconds duration is easily detectable in the unflagged data, but does not exceed the 0.125 Jy detection threshold in the image.

For fainter transients of $\lesssim3$ Jy, we also see flagging occurring.  This is because the connectivity of the data points making up the transient signal trigger the lower thresholds of the {\sc SumThreshold} algorithm.  Excluding the parts of Figure \ref{fig:11min-unflagged} where transients are too diluted to be detected, we find that $\sim10\%$ of transients are flagged due to this underlying connectivity.  This percentage was arrived at by taking the ratio of the number of surviving transients to the total number of injected transients.

\begin{figure}
\quad
\begin{subfigure}[b]{.5\textwidth}
\centering
\includegraphics[width=1\textwidth]{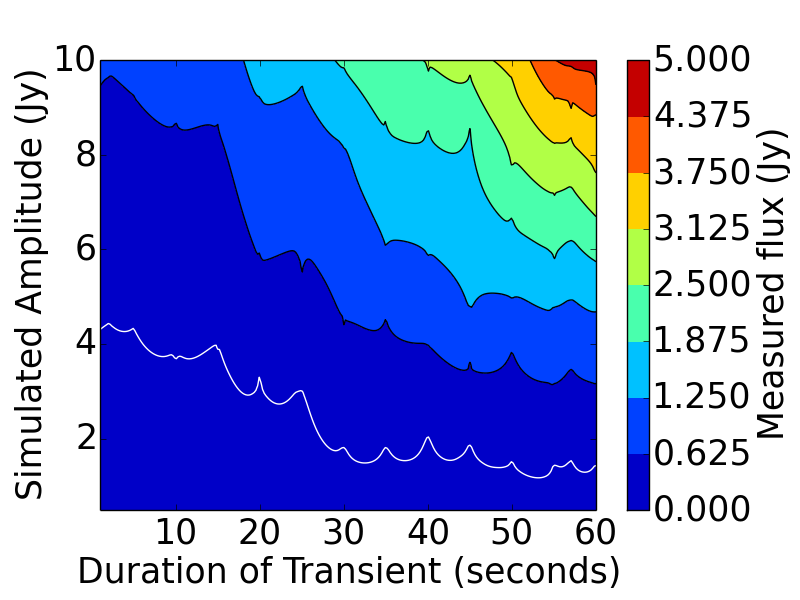}
\caption{Two minute measurement set with unflagged data.}
\label{fig:2min-unflagged}
\end{subfigure}
\quad
\begin{subfigure}[b]{.5\textwidth}
\centering
\includegraphics[width=1\textwidth]{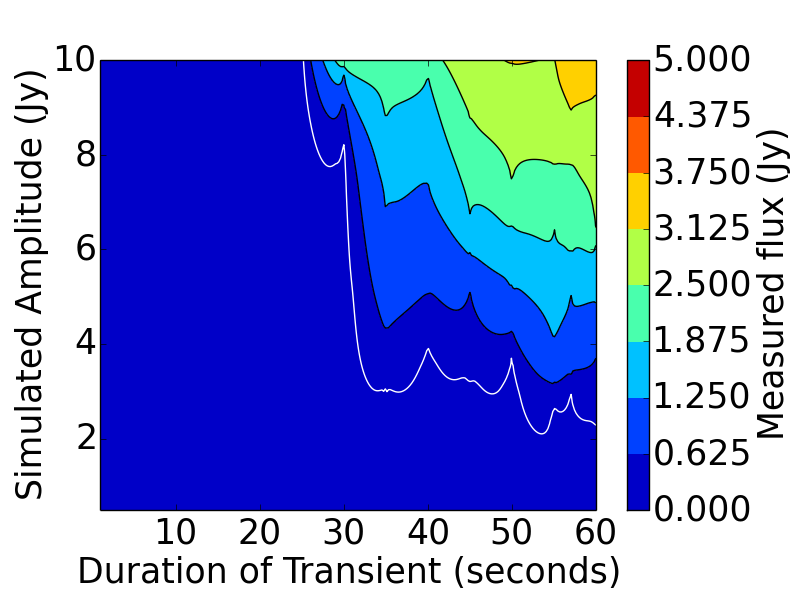}
\caption{Two minute data flagged with default flagger.}
\label{fig:2min-default}
\end{subfigure}
\quad
\begin{subfigure}[b]{.5\textwidth}
\centering
\includegraphics[width=1\textwidth]{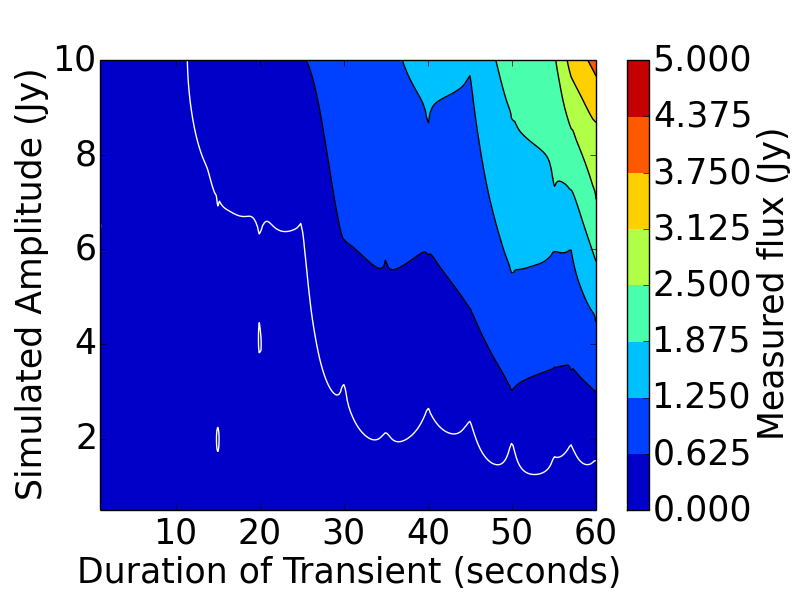}
\caption{Two minute flagged data with 6$\sigma$ time selection.}
\label{fig:2min-6sig}
\end{subfigure}
\quad
\begin{subfigure}[b]{.5\textwidth}
\centering
\includegraphics[width=1\textwidth]{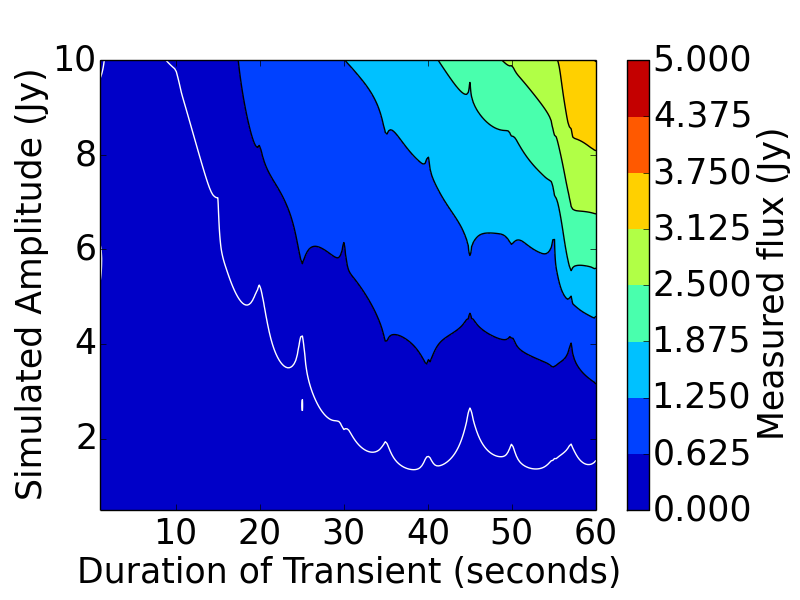}
\caption{Two minute flagged data, no time selection.}
\label{fig:2min-output}
\end{subfigure}
\quad
\caption{Observed fluxes for transient signals of a given amplitude and duration in a 2 minute measurement set.  Here, the top left plot is with unflagged data, the top right plot is for the default flagger, the bottom left is with the modified flagger with {\sc SumThreshold} only in the time direction but a time selection $\sigma >$ 6.5, and the bottom right is with {\sc SumThreshold} only in the time direction and with no time selection steps.  The white contour corresponds with $S_{min, 11min} = 0.55$ Jy, as measured from the RMS map as calculated by the sourcefinding script {\sc PySE}.}
  \label{fig:2minfluxes-mod}
\end{figure}

Similar effects are seen in the measurement sets of 2 minutes duration.  These results can be seen in Figure \ref{fig:2minfluxes-mod}, where again the upper left plot shows the fluxes measured in unflagged data, and the upper right plot shows results from the flagged data set.  We see a similar plot to what was seen before with the longer measurement set in that the briefest transients ($\simeq\Delta t < 25$ s) are all entirely flagged.  After this, transients of longer duration (25 s $< \Delta t < $60 s) are visible, but not at their full strengths because they are partially flagged in visibility space.  This can have consequences: for example, a 10 Jy, 20 second transient would be detectable in the unflagged data, but would not exceed the detection threshold in the flagged data.

Thus, flagging both template datasets results in loss of the brightest transients, as well as dilution of the recovered flux of unflagged transients.

\subsection{Modified Flagger Results}

In light of the results with default flagging, we carried out modifications of the flagger parameters, and tested the modified flagger on the same data.  Specifically, this test focused on two points: adjusting {\sc SumThreshold} to work ``in time direction" only, and to increase the $\sigma$ cutoff value for the time selection steps.  These tests were carried out for both the 2 minutes duration and 11 minutes duration measurement sets. 

First, we note that only adjusting one parameter but not the other did not yield a significant difference in flagging results compared with the default AOFlagger settings-- that is, if {\sc SumThreshold} was adjusted the transient would still be flagged as RFI by the $\sigma > 3.5$ time selection step, and vice versa.  When adjusting both {\sc SumThreshold} to only flag ``in time direction" and increasing the sigma cutoff for time selection, however, does yield results that are noticeably different from the default AOFlagger settings.

For the 11 minute measurement set, two such examples can be seen in the lower plots of Figure \ref{fig:11minfluxes}.  In both plots {\sc SumThreshold} is set to ``in time direction" only, but in the lower left panel (Figure \ref{fig:11min-6sig}) the time selection step is set to $\sigma > 6.5$, which we chose as an arbitrary cutoff to demonstrate the effect of a higher threshold for time selection.  In the lower right plot, Figure \ref{fig:11min-output}, the time selection is deleted from the algorithm altogether.  We see that although all the transients less than 60 seconds in duration are still flagged, there is some improvement in the number of transients that are visible because they are only partially flagged in both cases.  The scenario where the time selection steps are eliminated also shows better results than the panel with a $\sigma > 6.5$ time selection step included.  For example, while in Figure \ref{fig:11min-6sig} the $\lesssim3$ Jy signals are all flagged by the lower threshold, in Figure \ref{fig:11min-output} only the $\lesssim2.5$ Jy signals are now affected.  This can be attributed to the decoupling of the time dimension in the {\sc SumThreshold} connectivity based thresholding.  Further, the 4 Jy, 100 s transient described earlier which was flagged out would now exceed the detection threshold with the modified flagger that had no time selection present in the default AOFlagger, although would still be flagged if $6.5\sigma$ time selection was applied.

We see similar results for the 2 minute measurement set, where the modified flagger results are shown in the lower plots of Figure \ref{fig:2minfluxes-mod}, with the $\sigma >$ 6.5 cutoff on the left hand side plot, and eliminated altogether in the lower right plot.  Here, while all transients of a duration less than 20 seconds are still flagged, we have partial flagging occurring on a larger fraction of transients that are of (25 s $< \Delta t < $60 s) duration.  Once again, the lower right panel, which contains no time selection step, flags fewer transients than the lower left panel with a time selection step of $\sigma >$ 6.5.  Here, for example, we see that the 10 Jy, 20 second transient which did not exceed the detection limit with the default flagger now exceeds the detection limit in both modified flaggers.

\subsection{Observer Alert Script}
\label{sec:script}

\begin{figure}
\includegraphics[width=0.8\textwidth]{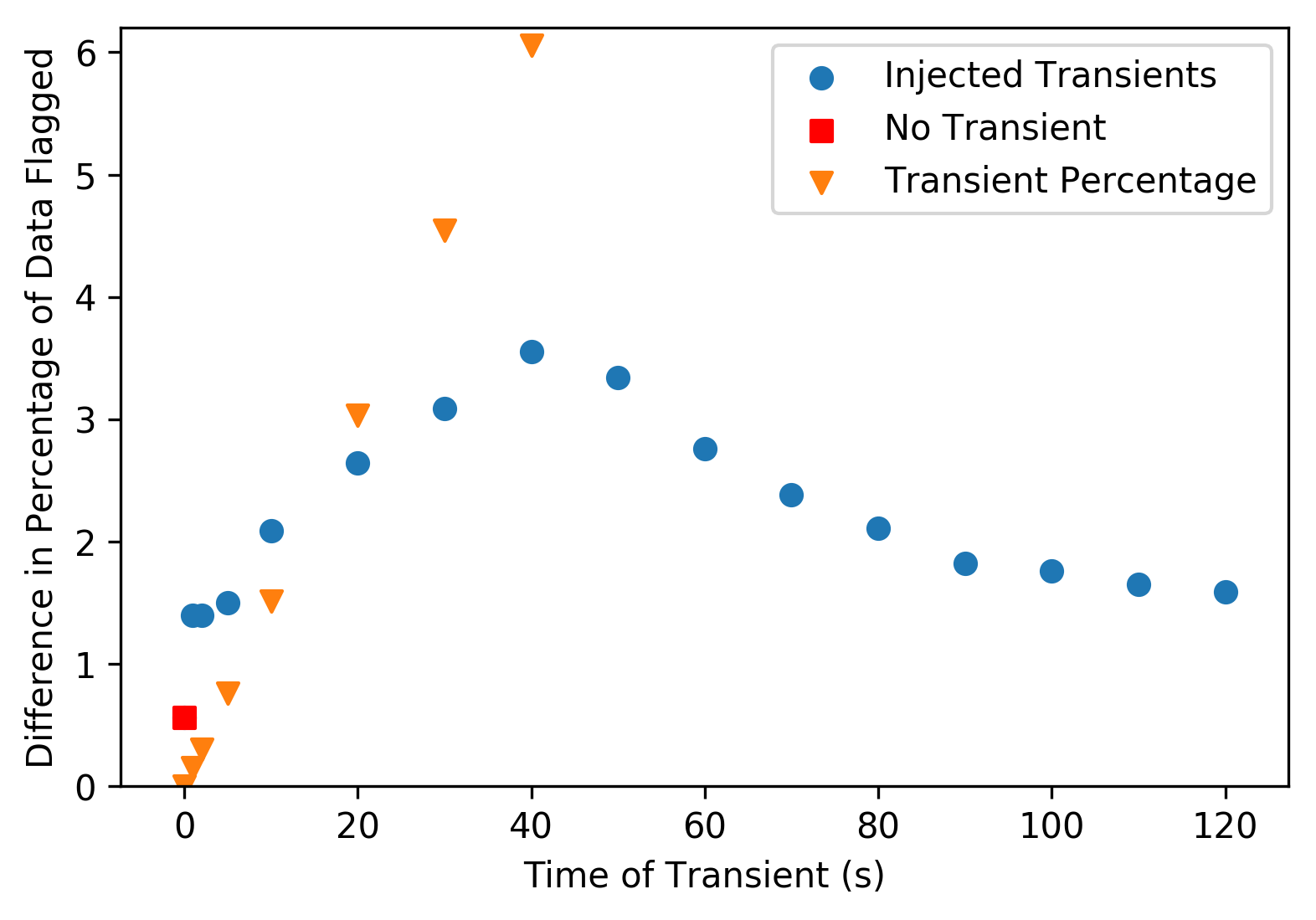}
\centering
\caption{The difference in percentage of data flagged between the default and modified strategies for an injected 10 Jy signal in an 11 minute measurement set, for different durations of time (blue).  We provide the difference for the same data set with no injected transient (red) for reference.  For the injected flagged transients, the difference between the two strategies steadily increases until it peaks at 40 seconds, and then begins to decrease (there is 1-4\% more data flagged by the default flagger, depending on the transient duration).  We attribute this decrease to the emergence of partial flagging, described in Section \ref{sec:default}, in transients of this duration and longer.  For reference, we also provide the percentage of the transient if it was completely flagged by the default but not at all by the modified flagger (yellow) until 40 seconds, when partial flagging begins to occur.  We will note that error bars in the average values in this plot were included, but are too small to be visible.}
\label{fig:percent-flagged}
\end{figure}

We have developed a script which can alert the observer when a potential transient is in the data stream.  This is based on a comparison on the percentages of data flagged between the two RFI flaggers, and various criteria which are expected to be different between transient RFI and an astrophysical signal.  In this script, first the RFI flagger checks the percentage of data flagged for the entire data set in each channel- a standard output given in AOFlagger- for both the default and modified strategies.  Although we expect the modified flagger to flag slightly less data than the default flagger, due to less stringent criteria, a transient in the data stream would also create a difference in the percentage flagged between the two methods.  As such, the difference between the percentage flagged can then be used to then determine whether a signal is in the data stream which may be worth further examination through further selection criteria.

To illustrate this phenomenon, we show the difference in percentages flagged for a 10 Jy signal of various durations injected into an 11 minute observation in Figure \ref{fig:percent-flagged}, illustrated with blue points.  We include the difference in percentage between the two strategies when no transient is injected in the data stream (red), and the percentage that would occur if the default flagger fully flagged the transient but the modified flagger did not (yellow), for reference.  What we find is even for the shortest duration transient we tested, of 1 second length, there is already a $\sim1.5\%$ difference between the percentage of the measurement set flagged between the default and modified flaggers, which exceeds the amount seen with no transient (where there is a 0.53\% difference).  Visual inspection of the data flags in rfigui confirmed this is because while both flaggers will flag these shortest transients in our case (Figure \ref{fig:11minfluxes}), the default flagger will additionally flag extra data before and after the transient which is not part of the signal.  The percentage flagged will then steadily increase as the percentage of time the signal is in the measurement set increases, until it peaks at 40 seconds, and begins to decrease.  This is the duration where partial flagging becomes a factor, where a fraction of the transient signal is flagged instead of the entire signal.  This results in the difference in percentages flagged beginning to decrease, although this difference still remains consistently higher than when there is no transient in the data.

Based on these tests, we decided that any difference in percentage of data flagged between the two strategies that exceeds 1\% is of interest for further inspection for these transients.  For this, we used {\sc pyrap} to sort the data by time series and to search for the following:

\begin{itemize}

\item First, we searched the data set for baselines where the columns FLAG\_ROW were set, which occurs when all the data in a row of visibilities are flagged.  We chose this parameter because a broadband transient signal is expected to be visible in all frequency channels in our data set, but a flag in only a few channels is expected to be RFI.  We should note that because LOFAR data sets often include a flagged row in the final time step of data, we set our script to disregard the last ten seconds of the measurement set.

\item Next, we searched the flagged row by baseline.  Specifically, we were interested in occurrences when FLAG\_ROW was true for a majority of baselines at a given time.  We chose this parameter because a flagged row only present in a small number of baselines is likely caused by local interference, whereas an astronomical signal would be visible in most baselines (although not all, depending on the noise within that specific baseline).

\item Finally, if all these criteria are met, the script returns a note to the observer, informing them a potential transient is present in their data.

\end{itemize}

We tested our script on our transient signals seen in Figure \ref{fig:percent-flagged}, which were 10 Jy signals in a 11 minute measurement set.  We found that when we used the above criteria, our script returned a potential transient alert for all our signals.  However, raising the difference in percentage flagged to 3\% only returned a potential transient alert for the transients of 30-50 second duration, and no alert was returned when the difference was set in excess of 4\%.

\section{Discussion}
\label{sec:discussion}

\subsection{Observed AOFlagger Effects}

From our simulation work, it appears that AOFlagger, which is relied on by many radio surveys searching for brief transient signals, could be flagging out existing bright, brief transients in the data stream.  In particular, it is possible all bright transients of just a few seconds' duration sufficiently bright to be detected in the integrated images have been flagged in previous surveys such as those of \citet{Stewart2016}, if they were present.  Re-examination of the data's flagged sections with the observer alert script could aid transient detection.  Using the observer alert script does not require the more computationally intensive imaging steps to identify previously flagged data which merits further inspection, but any astrophysical transient would originate from one point in the sky and converge to a point source upon imaging.  RFI, on the other hand, is much more likely to originate from the ground, meaning it would result in a noise-filled image if only affecting a few baselines.

Even when a transient is detected using default flagging methods, however, consideration should be given that the signal can be subject to partial flagging by RFI detection algorithms.  Further, some non-transient data is also flagged by the flagger, which can further affect measured fluxes.  Checking any candidate transient signals for partial flagging in the data stream (if possible) should be carried out in any transient detection follow up to determine the true flux density.

\subsection{Effect of the Duration of an Observation}

Further, it is important to consider the merits behind searching for very brief transients on the order of a few seconds in an 11 minute observation.  Although using such an observation length is not ideal for transients of this duration, often such observations are ``piggybacked" for transient searches from observations such as sky surveys.  For such a survey, even if imaging is done using shorter timescales than 11 minutes, the data are typically flagged over the full length of the measurement set before it is sliced into smaller increments for imaging in order to obtain a sufficiently large number of data points for flagging to be conducted.  This means that even if imaging is done in smaller increments that are more ideal for transient searches, potential transient signals on the order of a few seconds duration would be flagged.

In addition to this, our tests using a 2 minute observation show transients on the order of a few seconds duration are flagged even when flagging is conducted over a shorter time span when using LOFAR data.  There are surveys being conducted at these time scales, such as \citet{Rowlinson2016} which sought to constrain the transient rate by searching for signals on the order of a few seconds by taking 2 minute measurement sets from MWA and slicing them into 30 second images.  However, due to the differences between the LOFAR and MWA telescopes in SEFD and channel width, while the brightest transient signals would have been flagged, if they existed, it seems likely that fainter transient signals would have survived the flagging process.  Repeating the simulations in this work using an MWA measurement set would answer where precisely these cutoffs lie.  Further, if the brightest transient signals were flagged, the subsequent transient rate estimates from this and other surveys currently do not incorporate the effects of RFI flagging, and this should be taken into account.

\subsection{Effects of AOFlagger Modifications and Detection Script}

There are some modifications to AOFlagger that can be implemented in order to detect more transients than one would with the default settings (though, it should be emphasized, not all transients).  In particular, for AOFlagger, setting {\sc SumThreshold} to only operate in the time direction, and eliminating all time selection steps (i.e., only use frequency selection), would be advisable when seeking transients of $\sim$minute long durations within snapshots.  For transients on the order of tens of seconds long, flagging on a shorter time scale is advised, but consideration must be given to the number of time steps and resolution in the length of the data stream to ensure flagger accuracy.

We note that while these modifications to the RFI flagger will still result in automated flagging of narrow-band RFI, and not have significant effects on low-level RFI in general, implementing such a modified flagger opens up the possibility that bright, broadband RFI signals will no longer be flagged.

We have also presented in Section \ref{sec:script} a potential solution in which the differences between the default AOFlagger and modified version, along with the known distinguishing properties between RFI and transient signals, are used to alert the observer when a transient may be present in the data stream.  This script could be further modified depending on the class of transients in which the observer is interested- for example, a search for minutes-long transients such as that described in \citet{Stewart2016} could use a higher difference in percentage flagged than 1\%, which would decrease the number of false positives.  Further, the data could be examined for the sequential occurance of the script criteria in several time stamps, which would indicate a potential transient of longer duration.

We would recommend that an observer using this method choose their own parameters for the difference in percentages flagged, and the percentage of baselines with completely flagged visibilities at a given time, as these are dependent on the location of the telescope, and the resolution of the data acquired by the observer.  In particular, because some transient-like RFI will still meet the criteria outlined above, it is inevitable that a percentage of false positives will be present in the data stream.  We will investigate the effects of a modified flagger with alert script on a high-RFI observation with injected transients and different data resolutions in a future publication.

Another possibility could be to also implement the dictionary approach outlined by \citet{Czech2017}, which focused on filtering so-called transient RFI (defined in \citet{Czech2017} as non-constant RFI).  This work demonstrated the use of Markov models to identify transient RFI as a sequence of sub-events, and could reliably extract transient RFI events from a larger observation.  Given the similarities between transient RFI and astronomical sources, investigating the differences between the two categories with this method could provide another useful tool in distinguishing between the two.

\section{Conclusions}

Using the typical LOFAR imaging survey mode as an example, we have investigated the effects of automated RFI flagging on transient signals of astronomical origin.  We have demonstrated that RFI flagging with AOFlagger can flag out brief transients $< 60$ s in duration.  Further, we have calculated that  all bright transients $>10.6$ Jy would be flagged when the integration is 10 s and the frequency integration is 48 kHz.  Fainter signals of a few Jy brightness are also affected, depending on the extent of connectivity of the signal in the visibility plane.

For transient signals of $> 60$ s length in a measurement set of 11 minutes duration, the majority of the expected bright transient signals will survive the flagging process-- albeit with diminished flux due to ``partial flagging"-- and we have a good chance of observing these transients.  However, the original data would need to be reexamined in order to determine whether part of the signal was mistakenly flagged.  Reexamining the data may also reveal shorter transients already flagged in the data stream, which face a significant loss.  For a shorter LOFAR measurement set of just 2 minutes duration, it is equally likely that 100\% of radio transients of up to 20 seconds duration would be flagged.  Approximately 60\% of transients on the order of tens of seconds or longer will be detected, but partially flagged.

We have presented a script which can be used to alert an observer that the flagged data contains a potential astrophysical transient, which identified our simulated transients which had been flagged.  This could be used for any additional investigations into whether there are any transients in already flagged RFI data, in both previous survey data and in any future data preflagged to default settings.

For future transient surveys, in order to minimize the flagging of bright transients in the image plane we recommend modifying AOFlagger by only applying thresholding in the frequency direction, and eliminating the time selection steps in the algorithm that rely on sigma clipping.  With these steps in place, a larger fraction of bright transient signals will be seen, though likely partially flagged.  We also recommend the implementation of an observer alert system based on the differences between transient RFI and astronomical sources, to allow the correct identification of these signals.

\section*{Acknowledgements}

We acknowledge support from the European Research Council via the Advanced Investigator Grant no. 24729.  This work is also supported in part by European Research Council Advanced Grant 267697.  LOFAR, the Low Frequency Array designed and constructed by ASTRON, has facilities in several countries, that are owned by various parties (each with their own funding sources), and that are collectively operated by the International LOFAR Telescope (ILT) foundation under a joint scientific policy.  We would like to thank the LOFAR Observatory staff for their assistance in obtaining and the handling of this large data set.  We would also like to thank Andre Offringa for answering some initial questions.  S.C. acknowledges funding support from the UnivEarthS Labex program of Sorbonne Paris Cit\'e (ANR-10-LABX-0023 and ANR-11-IDEX-0005-02).




\begin{thebibliography}{40}

\bibitem[Adebahr et al.(2017)]{Adebahr2017} Adebahr, B., Krause, M., Klein, U., Heald, G., \& Dettmar, R.-J.\ 2017, AAP, 608, A29

\bibitem[Allison et al.(2017)]{Allison2017} Allison, J.~R., Moss, V.~A., Macquart, J.-P., et al.\ 2017, MNRAS, 465, 4450 

\bibitem[Barnbaum \& Bradley(1998)]{Barnbaum1998} Barnbaum, C., \& Bradley, R.~F.\ 1998, AJ, 116, 2598 

\bibitem[Bower et al.(2013)]{Bower2013} Bower, G.~C., Metzger, 
B.~D., Cenko, S.~B., Silverman, J.~M., \& Bloom, J.~S.\ 2013, ApJ, 763, 84 

\bibitem[Bowman et al.(2013)]{MWA} Bowman, J.~D., Cairns, 
I., Kaplan, D.~L., et al.\ 2013, PASA, 30, 31 

\bibitem[Broderick et al.(2016)]{Broderick2016} Broderick, J.~W., Fender, R.~P., Breton, R.~P., et al.\ 2016, MNRAS, 459, 2681 

\bibitem[Coenen et al.(2014)]{Coenen2014} Coenen, T., van Leeuwen, J., Hessels, J.~W.~T., et al.\ 2014, AAP, 570, A60 

\bibitem[Condon (1984)]{Cordon1984} Condon, J.~J.\ 1984, ApJ, 287, 461 

\bibitem[Czech et al.(2017)]{Czech2017} Czech, D., Mishra, A., \& Inggs, M.\ 2017, arXiv:1711.08823 

\bibitem[Fridman(2001)]{Fridman2001} Fridman, P.~A.\ 2001, AAP, 368, 369 

\bibitem[Fridman (2008)]{Fridman2008} Fridman, 2008, AJ, 35, 1810

\bibitem[van Haarlem et 
al.(2013)]{LOFAR} van Haarlem, M.~P., Wise, M.~W., Gunst, A.~W., et al.\ 2013, A\&A, 556, A2 

\bibitem[Heald et al.(2011)]{Heald2011} Heald, G., Bell, M.~R., 
Horneffer, A., et al.\ 2011, Journal of Astrophysics and Astronomy, 32, 589 

\bibitem[Hyman et al.(2009)]{Hyman2009} Hyman, S.~D., Wijnands, R., Lazio, T.~J.~W., et al.\ 2009, ApJ, 696, 280 

\bibitem[Jaeger et al.(2012)]{Jaeger2012} Jaeger, T.~R., Hyman, S.~D., Kassim, N.~E., \& Lazio, T.~J.~W.\ 2012, AJ, 143, 96 

\bibitem[Kocz et al.(2012)]{Kocz2012} Kocz, J., Bailes, M., Barnes, D., Burke-Spolaor, S., \& Levin, L.\ 2012, MNRAS, 420, 271 

\bibitem[Lorimer et al.(2007)]{Lorimer2007} Lorimer, D.~R., Bailes, M., McLaughlin, M.~A., Narkevic, D.~J., \& Crawford, F.\ 2007, Science, 318, 777 

\bibitem[Loose(2008)]{Loose2008} Loose, G.~M.\ 2008, Astronomical 
Data Analysis Software and Systems XVII, 394, 91 

\bibitem[McKean et al.(2011)]{McKean2011} McKean, J., Ker, L., van Weeren, R.~J., et al.\ 2011, arXiv:1106.1041 

\bibitem[Nita \& Gary(2010)]{Nita2010} Nita, G.~M., \& Gary, D.~E.\ 2010, PASP, 122, 595 

\bibitem[Nita(2016)]{Nita2016} Nita, G.~M.\ 2016, MNRAS, 458, 2530 

\bibitem[Offringa(2012)]{Offringathesis} Offringa, A.~R.\ 2012, 
Ph.D.~Thesis, University of Groningen  

\bibitem[Offringa et 
al.(2012)]{AOFlagger} Offringa, A.~R., van de Gronde, J.~J., \& Roerdink, J.~B.~T.~M.\ 2012,  A\&A, 539, AA95 

\bibitem[Offringa et al.(2012)]{Offringa2012} Offringa, A.~R., de Bruyn, A.~G., \& Zaroubi, S.\ 2012, MNRAS, 422, 563 

\bibitem[Offringa et al.(2010)]{SumThreshold} Offringa, A.~R., de 
Bruyn, A.~G., Biehl, M., et al.\ 2010, MNRAS, 405, 155 

\bibitem[Offringa et al.(2010)]{LOFAR-AOFlagger} Offringa, A.~R., de Bruyn, A.~G., Zaroubi, S., \& Biehl, M.\ 2010, arXiv:1007.2089 

\bibitem[Offringa et al.(2013)]{LOFAR-RFI} Offringa, A.~R., de Bruyn, A.~G., Zaroubi, S., et al.\ 2013,  A\&A, 549, A11 

\bibitem[Offringa et al.(2015)]{OffringaMWA} Offringa, A.~R., 
Wayth, R.~B., Hurley-Walker, N., et al.\ 2015, PASA, 32, e008 

\bibitem[Price et al.(2017)]{Price2017} Price, D.~C., Greenhill, L.~J., Fialkov, A., et al.\ 2017, arXiv:1709.09313 

\bibitem[Rane \& Lorimer(2017)]{Rane2017} Rane, A., \& Lorimer, D.\ 2017, Journal of Astrophysics and Astronomy, 38, 55 

\bibitem[Raza et al.(2002)]{Raza2002} Raza, J., Boonstra, A.-J., \& van der Veen, A.~J.\ 2002, IEEE Signal Processing Letters, 9, 64 

\bibitem[Rowlinson et al.(2016)]{Rowlinson2016} Rowlinson, A., Bell, M.~E., Murphy, T., et al.\ 2016, MNRAS, 458, 3506 

\bibitem[Ryabov et al.(2004)]{Ryabov2004} Ryabov, V.~B., Zarka, P., \& Ryabov, B.~P.\ 2004, PLANSS, 52, 1479 

\bibitem[Smol{\v c}i{\'c} et al.(2016)]{ATCA2016} Smol{\v c}i{\'c}, V., Delhaize, J., Huynh, M., et al.\ 2016, AAP, 592, A10

\bibitem[Stewart et al.(2016)]{Stewart2016} Stewart, A.~J., Fender, R.~P., Broderick, J.~W., et al.\ 2016, MNRAS, 456, 2321 

\bibitem[Swinbank et al.(2015)]{Swinbank2015} Swinbank, J.~D., Staley, T.~D., Molenaar, G.~J., et al.\ 2015, Astronomy and Computing, 11, 25 

\bibitem[Taylor et al.(1999)]{Taylor1999} Taylor, G.~B., Carilli, C.~L., \& Perley, R.~A.\ 1999, Synthesis Imaging in Radio Astronomy II, 180, Chapter 9.

\bibitem[Tasse et 
al.(2013)]{Tasse2013} Tasse, C., van der Tol, S., van Zwieten, J., van Diepen, G., \& Bhatnagar, S.\ 2013,  A\&A, 553, A105 

\bibitem[Tingay et al.(2013)]{Tingay2013} Tingay, S.~J., Goeke, R., Bowman, J.~D., et al.\ 2013, PASA, 30, e007

\bibitem[Tingay et al.(2015)]{Tingay2015} Tingay, S.~J., Trott, 
C.~M., Wayth, R.~B., et al.\ 2015, AJ, 150, 199 

\bibitem[Trott et al.(2013)]{Trott2013} Trott, C.~M., Tingay, S.~J., \& Wayth, R.~B.\ 2013, ApJL, 776, L16 

\bibitem[van Velzen et al.(2013)]{vanVelzen2013} van Velzen, S., Frail, D.~A., K{\"o}rding, E., \& Falcke, H.\ 2013, AAP, 552, A5 



\bibitem[Williams et al.(2017)]{Williams2017} Williams, P.~K.~G., Gizis, J.~E., \& Berger, E.\ 2017, APJ, 834, 117 


\end{thebibliography}


\end{document}